\documentclass[aps,prb,twocolumn,superscriptaddress,10pt]{revtex4-2}

\usepackage[utf8]{inputenc}
\usepackage{bbm}
\usepackage[T1]{fontenc}
\usepackage{amsmath,amssymb}
\usepackage{graphicx}
\usepackage{hyperref}
\usepackage{xcolor}
\usepackage{physics} 
\usepackage{siunitx}  
\usepackage[normalem]{ulem}

\hypersetup{
    colorlinks=true,
    linkcolor=blue,
    citecolor=blue,
    urlcolor=blue
}

\begin{document}

\title{
Real-space first-principles approach to orbitronic phenomena in metallic multilayers}

\author{Ramon Cardias}
\email{ramon.cardias@cbpf.br}
\affiliation{Centro Brasileiro de Pesquisas Físicas (CBPF), Rua Dr Xavier Sigaud 150, Urca, 22290-180, Rio de Janeiro-RJ, Brazil}

\author{Hugo U. R. Strand}
\affiliation{
School of Science and Technology, \"Orebro University, SE-70182 Örebro,
Sweden
}

\author{Anders Bergman}
\affiliation{
Department of Physics and Astronomy, Uppsala University, Box 516,
SE-75120 Uppsala, Sweden
}

\author{A. B. Klautau}
\affiliation{Faculdade de F\'\i sica, Universidade Federal do Par\'a, Bel\'em, PA, Brazil}

\author{Tatiana G. Rappoport}
\affiliation{Centro Brasileiro de Pesquisas Físicas (CBPF), Rua Dr Xavier Sigaud 150, Urca, 22290-180, Rio de Janeiro-RJ, Brazil}
\affiliation{International Iberian Nanotechnology Laboratory (INL), Av. Mestre José Veiga, 4715-330 Braga, Portugal}

\date{\today}

\begin{abstract}
We develop a real-space first-principles method based on density functional theory to investigate orbitronic phenomena in complex materials. Using the Real-Space Linear Muffin-Tin Orbital method within the Atomic Sphere Approximation (RS-LMTO-ASA) combined with a Chebyshev polynomial expansion of the Green’s functions, we compute orbital (spin) Hall transport and orbital (spin) accumulation directly in real space.  The approach scales linearly with system size and naturally incorporates disorder, finite-size effects, and interface roughness. We apply the method to transition-metal-based heterostructures and demonstrate the emergence of substantial orbital (spin) accumulation, even in centrosymmetric systems. Our methodology provides a scalable and flexible framework for realistic simulations of orbital transport phenomena in complex heterostructures.
\end{abstract}

\maketitle

\section{Introduction}
The study of orbitronics, which exploits the orbital degree of freedom of electrons \cite{Go2021a, Jo2024, Wang2024} for information processing and spintronic applications, has garnered growing attention in recent years. In contrast to spin-based approaches, orbitronic phenomena offer new functionalities arising from the intrinsic orbital structure of electronic states, including the generation of orbital angular momentum (OAM) currents, orbital accumulation at interfaces \cite{Yoda2018, Johansson2024, Nikolaev2024, Chirolli2022,Sala2022},  orbital torque \cite{Go2020a, Go2020b, Bose2023, Fukunaga2023, Santos2023, Lyalin2024} and pumping mechanisms \cite{Go2023,Han2023a,Hayashi2024,Santos2024,Wang2025} mediated by orbital degrees of freedom. Among these, the orbital Hall effect (OHE) – the transverse flow of orbital angular momentum under an electric field \cite{Bernevig2005,Kontani2008,Kontani2009, Tanaka2008,Go2018,Salemi2022,Choi2023,Lyalin2023, Sala2022,Sala2023,Abrao2025} 
– has emerged as a particularly promising mechanism for generating orbital torques in systems where strong spin-orbit coupling converts orbital flow into spin accumulation \cite{Go2020a, Go2020b, Bose2023, Fukunaga2023, Santos2023, Lyalin2024}.

While considerable progress has been made in understanding orbitronic effects in quantum bulk~\cite{Busch2023,Yen2024,Brinkman2024,Atencia2024,Liu2025,Hagiwara2025} and 2D materials \cite{Canonico2020a, Canonico2020b, Bhowal2020,Bhowal2020a,Bhowal2021,Cysne2021a, Cysne2022,Costa2023, Pezo2023,Chen2024,Veneri2025,Faridi2025,Sun2025}, transition metals (TMs) remain a central platform for both fundamental investigations and technological applications. Their complex d-band structure and strong hybridization effects pose theoretical challenges but also enable substantial orbital currents, as demonstrated in recent experimental and computational studies \cite{Kontani2008,Kontani2009, Tanaka2008,Go2018,Salemi2022,Choi2023,Lyalin2023, Sala2022,Sala2023,Abrao2025,Go2020a, Go2020b, Bose2023, Fukunaga2023, Santos2023, Lyalin2024,Go2023,Han2023a,Hayashi2024,Santos2024,Wang2025}. In particular, metallic multilayer structures involving TMs provide a controllable way to probe orbital accumulation, injection, and torque at interfaces.

In this work, we develop a real-space density functional theory (DFT) approach based on the Real-Space Linear Muffin-Tin Orbital method within the Atomic Sphere Approximation (RS-LMTO-ASA) \cite{frota-pessoa-1992,klautauMagneticPropertiesCo2004,PhysRevLett.71.4206,klautauOrbitalMoments3d2005} to investigate orbitronic phenomena in transition-metal multilayers. In contrast to standard approaches that begin with momentum-space DFT calculations and subsequently construct tight-binding Hamiltonians via Wannier \cite{Salemi2022,Go2024} or pseudo-atomic orbital projections \cite{Costa2023}, we retain the full real-space DFT Hamiltonian. This allows us to compute transport quantities such as orbital accumulation and the orbital Hall effect directly in real space. For this purpose, we employ a Chebyshev polynomial expansion of the Green’s functions to evaluate the Kubo-Bastin formula in linear response \cite{Garcia2015,Garcia2016,Ferreira2015,Cysne2016,Joao2020, Pires2022,Castro2024}. Our method follows the strategy of large-scale real-space quantum transport calculations as implemented in open-source packages such as Kwant \cite{Groth_2014} and Kite \cite{Joao2020}, but here it is fully integrated with a real-space first-principles description. 

A key advantage of our approach is its scalability and flexibility. Both RS-LMTO-ASA and Chebyshev expansion method scale linearly with the number of nonequivalent atoms in the system, allowing for the treatment of large-scale heterostructures and multilayers with complex geometries. Since all calculations are performed directly in real space, the method naturally accommodates open boundary conditions and enables the inclusion of structural imperfections, compositional disorder, and interface roughness without additional approximations. This makes it particularly well-suited for simulating realistic systems relevant to orbitronic phenomena, where spatial inhomogeneity and finite-size effects play a crucial role.

We focus on orbital (spin) Hall transport and non-equilibrium orbital (spin) accumulation in layered TM systems an a series of TM-based heterostructures. Our results demonstrate that substantial orbital (spin) accumulation can arise even in centrosymmetric systems, driven purely by band-structure asymmetries and interfacial scattering, and that the interplay between OHE and structural confinement can be used to engineer torque generation in TM bilayers.

\section{Methodology}
This section is divided into two parts. In the first part, we describe the real-space first-principles calculations based on the RS-LMTO-ASA method, which provide the electronic structure and real-space Hamiltonians of the transition-metal multilayers. In the second part, we detail the linear response formalism used to compute the relevant transport quantities, such as orbital accumulation and orbital Hall conductivity, directly from the real-space Hamiltonian using a Chebyshev polynomial expansion of the Green’s functions.
\subsection{First-principle calculations}
The RS-LMTO-ASA \cite{frota-pessoa-1992,klautauMagneticPropertiesCo2004,PhysRevLett.71.4206,Kvashnin2016,bergmanMagneticInteractionsMn2006,bergmanMagneticStructuresSmall2007,cardias2020first,szilvaTheoryNoncollinearInteractions2017,Brandao2025,klautauOrbitalMoments3d2005,PhysRevB.103.L220405,PhysRevB.85.014436,rodriguesFinitetemperatureInteratomicExchange2016,PhysRevB.83.014406} is a first-principles, 
self-consistent electronic structure method based on the 
LMTO-ASA formalism \cite{andersenLinearMethodsBand1975}, and uses the recursion method to solve the eigenvalue problem directly in real space.

It is particularly suited for studying systems with 
complex geometries and lacking translational symmetry, such as 
surfaces, multilayers, substitutional and interstitial 
impurities, and clusters embedded in these hosts, as well as 
nanostructured materials \cite{carvalho2023correlation,PhysRevB.108.224408,PhysRevB.94.014413,PhysRevB.105.224413,PhysRevB.93.014438,PhysRevB.60.3421,bezerra-netoComplexMagneticStructure2013,cardiasMagneticElectronicStructure2016,frota-pessoaInfluenceInterfaceMixing2002}. Also, this method has been generalized to describe noncollinear magnetism \cite{bergmanMagneticInteractionsMn2006,bergmanMagneticStructuresSmall2007,cardias2020first,szilvaTheoryNoncollinearInteractions2017,PhysRevB.108.224408}. 

It is a method that scales linearly with the number of nonequivalent atoms, and the solutions are accurate 
around a given energy $E_\nu$,  usually taken as the center
of gravity of the occupied part of the $s$, $p$, and $d$ bands 
being considered. We work in the orthogonal representation of the LMTO-ASA 
formalism, and expand the orthogonal 
Hamiltonian in terms of tight-binding (TB) parameters, 
for the better use of the recursion method. 
In this context, the resulting Hamiltonian can be written as
 \begin{equation}
H = E_\nu + \bar{h} + \bar{h} \bar{o} \bar{h},
\end{equation}
where $\bar{h}$  is a Hermitian matrix, given 
by $\bar{h} = \bar{C} - E_\nu + \bar{\Delta}^{1/2} \bar{S} \bar{\Delta}^{1/2}$. 
The quantities $\bar{C}$, $\bar{\Delta}$, $\bar{o}$ are potential parameters, 
while $\bar{S}$ denotes the structure constants 
in the tight-binding LMTO-ASA representation. For details see Ref. \onlinecite{CardiasPhD2018}.

The RS-LMTO-ASA self-consistency process is achieved through iterative updates of the electronic occupation, recalculation of potential parameters, and solution of the eigenvalue problem \cite{frota-pessoa-1992}. An essential step in this process involves accurately calculating the Local Density of States (LDOS), which is critical for determining the band moments and facilitating convergence. As mentioned, to solve the eigenvalue problem in real space and find the LDOS 
at each nonequivalent site, we use the recursion method.

Traditionally, the Lanczos recursion algorithm has been used to tridiagonalize the Hamiltonian, enabling the LDOS reconstruction via continued fraction expansion \cite{haydockRecursiveSolutionSchrodinger1980,beerRecursionMethodEstimation1984}. However, the Lanczos method can suffer from numerical instabilities, especially problematic when treating systems with substantial $s$- and $p$-orbital contributions, such as insulators and oxides.

To overcome these limitations, a Chebyshev polynomial-based recursion scheme has been implemented in the RS-LMTO-ASA method. In this approach, the LDOS is computed in terms of a Chebyshev polynomial expansion :
\begin{equation}
f(x) = \frac{1}{\pi \sqrt{1 - x^2}} \left[ \mu_0 + 2 \sum_{n=1}^\infty \mu_n T_n(x) \right],
\label{eq:chebdos}
\end{equation}
where $\mu_n = \int_{-1}^{1} f(x) T_n(x) \, dx$ are the Chebyshev moments and $T_n(x)$ are the Chebyshev polynomials of the first kind. 

The Chebyshev moments  can be computed as:
$\mu_m = \langle \beta | T_n(\tilde{H}) | \alpha \rangle$, 
where $|\alpha\rangle$ and $|\beta\rangle$ are states of the system. Given an initial state $|\alpha\rangle$, it is possible to  iteratively construct the states $|\alpha_n\rangle = T_n(\tilde{H}) |\alpha\rangle$ using the following recurrence relations:
$|\alpha_0\rangle = |\alpha\rangle$,  
$|\alpha_1\rangle = \tilde{H} |\alpha_0\rangle$,  and
$|\alpha_{n+1}\rangle = 2 \tilde{H} |\alpha_n\rangle - |\alpha_{n-1}\rangle$, with $\tilde{H} = \frac{H - b}{a}$ being the rescaled Hamiltonian to ensure its spectrum lies within $[-1, 1]$.
The Chebyshev moments are then obtained by projecting onto $|\beta\rangle$:
$\mu_n = \langle \beta | \alpha_n \rangle$. For the LDOS, we have $| \alpha \rangle = | \beta \rangle$, which is the identity matrix $\mathbbm{1}$ at the nonequivalent site $i$. The total density of states will then be the sum of the LDOS, i.e. one recursion procedure per nonequivalent atom is needed to obtain the moments $\mu_n$, hence the linear scaling.



The rescaling parameters $a$ and $b$ defining $\tilde{H}$ are determined from the extreme eigenvalues of the Hamiltonian:
\begin{equation}
a = \frac{E_{\text{max}} - E_{\text{min}}}{2 - \epsilon}, \quad b = \frac{E_{\text{max}} + E_{\text{min}}}{2},
\end{equation}
where $\epsilon$ is a small positive constant introduced to avoid numerical instabilities.

To improve convergence, the Kernel Polynomial Method (KPM) \cite{weise_kernel_2006} is employed, introducing a damping factor $g_n$ (specifically the Jackson kernel) to the moments, effectively truncating the infinite series and yielding a smooth and stable LDOS. For the systems studied in this paper, we truncated the expansion in Eq.~\ref{eq:chebdos} with $N_n = 500$ Chebyshev moments.

The Chebyshev recursion allows for robust and efficient computation of the LDOS, especially for systems where accurate description of $s$- and $p$-orbital behavior and band gaps is crucial. Unlike the Lanczos method, the Chebyshev-based approach maintains numerical stability even for a large number of recursion steps, thereby avoiding the generation of spurious states and improving the reliability of the electronic structure description.

\subsection{Linear-response}
An important advantage of working with a real-space Hamiltonian directly obtained from first-principles calculations is that it eliminates the need for intermediate projection schemes, such as maximally localized Wannier functions or pseudo-atomic orbitals, which are commonly used to construct tight-binding models from momentum-space DFT calculations. This direct access to the full real-space Hamiltonian makes the method particularly well-suited for linear response calculations formulated entirely in real space. In particular, we employ a Chebyshev polynomial expansion of the Green’s functions to evaluate transport coefficients efficiently and systematically. This approach builds upon the same polynomial expansion already used to compute the local density of states, ensuring consistency across spectral and transport quantities while preserving the full complexity of the first-principles electronic structure.

To compute transport quantities such as spin or orbital conductivities, we use the Kubo-Bastin formalism, which expresses the linear response coefficients in terms of Green’s functions and current operators\cite{Garcia2015,Garcia2016,Ferreira2015,Cysne2016,Joao2020, Pires2022,Castro2024}.

We define a generalized current operator \( \hat{J}_\alpha^{\hat{\mathcal{O}_j}} \), where \( \alpha \) denotes the spatial direction, $\hat{\mathcal{O}}$ is $\hat{S}$ or $\hat{L}$ and \( j\in\{x,y,z\} \) indicates the orientation of the spin/OAM operator \cite{Garcia2016}. The input operator \( \hat{J}_\beta^{c} = -e \hat{v}_\beta \) corresponds to the charge current driven by the electric field \( E^\beta \), while \( \hat{J}^{i,\hat{\mathcal{O}}}_{\alpha} \) denotes the output current of interest. The spin and orbital currents take the form \( \hat{J}_\alpha^{\hat{\mathcal{O}}_j} = \frac{1}{2} \{ \hat{v}_\beta,  \hat{\mathcal{O}}_j \} \), with \( \hat{\mathcal{O}}_j = \hat{S}_j \) or \( \hat{L}_j \), respectively.

The linear response of \( \hat{J}^{\mathcal{O}_j}_\alpha \) to an applied electric field in direction \( \beta \) is computed using the Kubo-Bastin formula \cite{Garcia2015,Garcia2016,Ferreira2015,Cysne2016,Joao2020, Pires2022,Castro2024}:
\begin{widetext}
\begin{equation}
\sigma_{\alpha\beta}^{\hat{\mathcal{O}}_j}(\mu, T) = \frac{\hbar}{\pi \Omega} \int dE\, f(E, \mu, T)\, 
\Re\, \mathrm{Tr} \left[\text{i} \hat{J}^{\hat{\mathcal{O}}_j}_\alpha \frac{d\hat{G}(E)}{dE} \hat{J}_{\beta}^c \, \Im \hat{G}(E) \right],
\label{eq:kubo-bastin-general}
\end{equation}
\end{widetext}
where \( \hat{G}(E) = (E + \text{i} \eta - \hat{H})^{-1} \) is the retarded Green’s function, \( f(E,\mu,T) \) is the Fermi-Dirac distribution for a given temperature T and chemical potential $\mu$, \( \hat{H} \) is the real-space Hamiltonian obtained from first principles, and \( \Omega \) is the system volume. 

 To evaluate this expression numerically, we employ a Chebyshev expansion of the Green’s functions. The Hamiltonian is rescaled to the interval \( [-1, 1] \), and all energies are mapped to the dimensionless variable \( \varepsilon \). The conductivity tensor at site \( i \) is then given by

\begin{equation}
\sigma_{\alpha\beta}^{j,\hat{\mathcal{O}}}(i; \mu, T) \approx \frac{i \hbar}{4 \Omega} \int_{-1}^{1} d\tilde{\varepsilon}\, f(\tilde{\varepsilon}, \tilde{\mu}, \tilde{T})\, 
\tilde{\zeta}^{\alpha\beta,\hat{\mathcal{O}}}_i(\tilde{\varepsilon}),
\end{equation}

with
\begin{equation}
\tilde{\zeta}^{\alpha\beta,\hat{\mathcal{O}}}_i(\varepsilon) = \sum_{m,n=0}^{M-1} g_{mn}(\tilde{\varepsilon})\, \mu_{mn}^{\alpha\beta,\hat{\mathcal{O}}}(i),
\end{equation}
\begin{equation}
\mu_{mn}^{\alpha\beta,\hat{\mathcal{O}}}(i) = \left\langle i \middle| T_m(\tilde{H}) \hat{J}^{\alpha,\hat{\mathcal{O}}} T_n(\tilde{H}) \hat{J}^{\beta,c} \middle| i \right\rangle.
\end{equation}

Here, \( T_n(\tilde{H}) \) denotes the \( n \)-th Chebyshev polynomial of the rescaled Hamiltonian, and \( g_n \) are kernel damping coefficients that regularize the truncated series. The energy-dependent prefactor $g_{mn}$ reads \cite{Garcia2015}
\begin{multline}
g_{mn}(\tilde{\varepsilon}) = g_m g_n (1 - \tilde{\varepsilon}^2)^{-2} \times \\
[ (\tilde{\varepsilon} - \text{i} n \sqrt{1 - \tilde{\varepsilon}^2}) e^{\text{i} n \arccos \tilde{\varepsilon}} T_m(\tilde{\varepsilon})\\
+ (\tilde{\varepsilon} + \text{i} m \sqrt{1 - \tilde{\varepsilon}^2}) e^{-\text{i} m \arccos \tilde{\varepsilon}} T_n(\tilde{\varepsilon}) ].
\end{multline}

This formulation allows for the evaluation of spin or orbital transport coefficients in systems with arbitrary complexity, including real-space Hamiltonians.

All required quantities are evaluated directly in real space using a Chebyshev polynomial expansion, consistent with the procedure previously employed for calculating the local density of states. Instead of relying on stochastic trace evaluations, commonly used in large-scale real-space methods, we compute observables locally at a representative site \( i \). This is made possible by exploiting translational symmetry in the directions where the system is periodic, while the Chebyshev recursion naturally incorporates the contributions from all other sites.

To ensure convergence, we increase the lateral dimensions of the simulated region until the calculated quantities become independent of the system size \( L \). In the case of bulk materials, we typically use supercells containing 21 unit cells in each spatial direction and evaluate the response at a site \( i \) located at the center of the system, taking advantage of translational invariance in all three directions. For multilayers and heterostructures, where symmetry is broken along the out-of-plane direction, we retain translation symmetry in the \( x \)–\( y \) plane and select one representative site \( i \) per layer, located at the center of the layer. This strategy provides direct access to the spatial profile of current responses across the heterostructure with atomic resolution. Here, we used the same number of Chebyshev moments $N_n = 500$ to maintain consistency across both self-consistent and transport calculations. For the kernel damping coefficients $g_m$, we used the Lorentz kernel~\cite{weise_kernel_2006}, 
%
%
\begin{equation}
    g_{n}^{L}=\frac{\sinh [\lambda(1-n / N)]}{\sinh (\lambda)} .
    \label{eq:lorentzkernel}
\end{equation}
since it allows us to set the parameter $\lambda$ which will work as a broadening parameter, as opposed to the Jackson kernel, where the broadening is intrinsically connected with the number of the moments in the expansion and cannot be fixed.

The same Chebyshev-based real-space methodology is employed to compute spin and orbital accumulations. In this case, instead of evaluating current operators, we calculate the linear response of a generalized observable \( \hat{\mathcal{O}} \) to an applied electric field. For spin or orbital accumulation, \( \hat{\mathcal{O}}_j \) corresponds to the spin or orbital angular momentum operator \( \hat{S}_j \) or \( \hat{L}_j \). The accumulation response is then captured by the magnetoelectric (ME) susceptibility tensor \( \chi_{j\beta}^{\hat{\mathcal{O}}} \)  which quantifies the induced moment in direction \( j\in\{x,y,z\} \) per unit electric field in direction \( \beta \). Therefore, both conductivities $\sigma_{\alpha\beta}^{\hat{\mathcal{O}}_j}$ and ME susceptibility tensors \( \chi_{j\beta}^{\hat{\mathcal{O}}} \) can be written, respectively, in a linear-response formalism as follows \cite{Salemi2021}

\begin{equation}
    \begin{aligned}
\delta \hat{\mathcal{O}}_j & =\chi_{j \beta}^{\hat{\mathcal{O}}} E^{\beta}, \\
J_{\alpha}^{\hat{\mathcal{O}}_j} & =\sigma_{\alpha \beta}^{\hat{\mathcal{O}}_j}E^{\beta},
\end{aligned}
\end{equation}

This unified framework allows us to compute not only current responses, but also field-induced observables with detailed layer resolution,  relevant to spintronics and orbitronics.

\begin{figure*}[ht]
    \includegraphics[width=2\columnwidth]{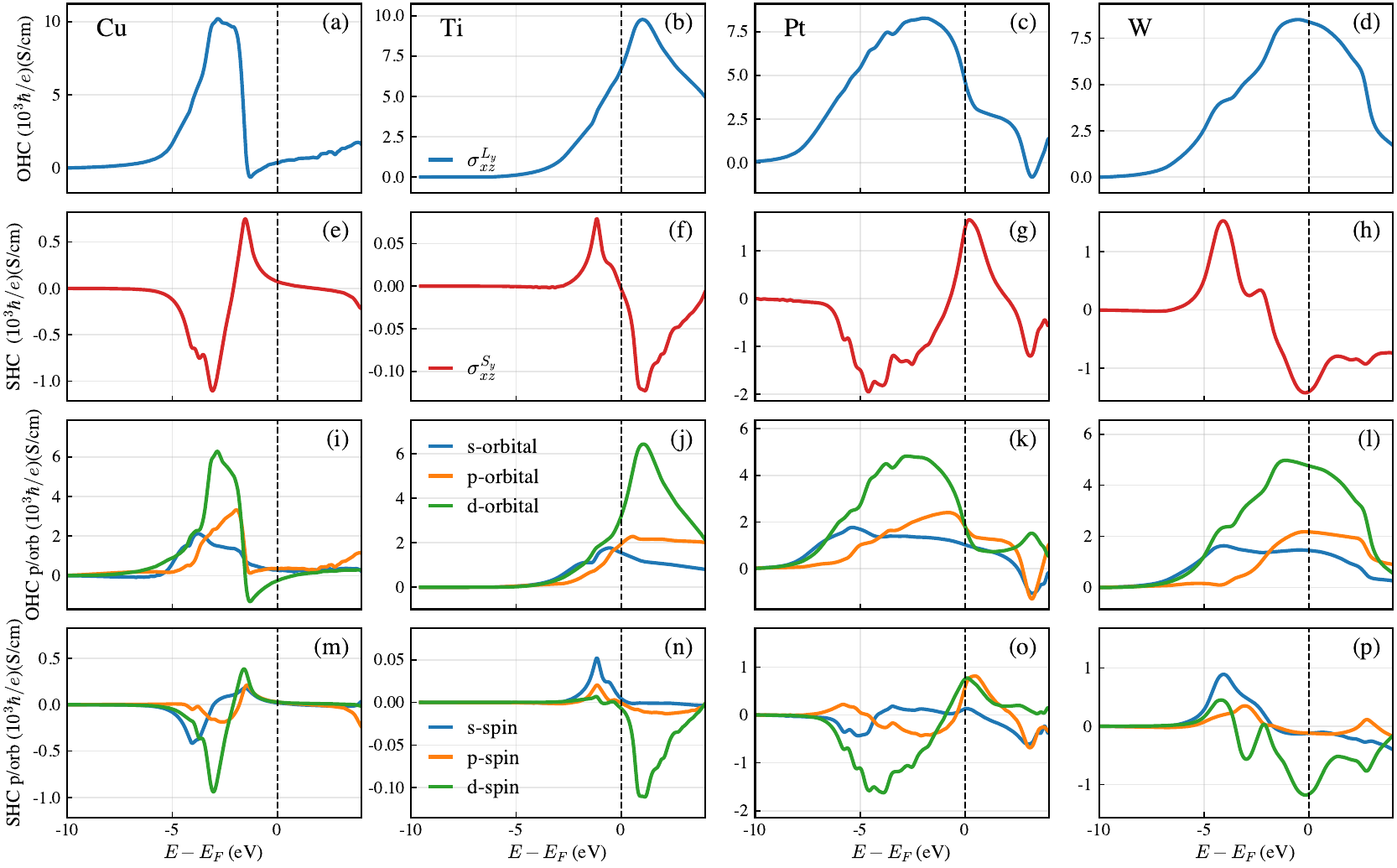}
    
    \caption{
 Orbital and spin Hall conductivities of selected nonmagnetic transition metals. The top two rows display the orbital Hall conductivity (OHC) (top row) and spin Hall conductivity (SHC) (second row). The bottom two rows show the orbital-resolved decomposition of OHC (third row) and SHC (fourth row) into $s$-, $p$-, and $d$-orbital contributions. Columns correspond to different metals: Cu (first column: a,e,i,m), Ti (second column: b,f,j,n), Pt (third column: c,g,k,o), and W (fourth column: d,h,l,p). 
   }
    \label{fig:condnm}
\end{figure*}
\section{Spin and Orbital Conductivities}

We begin by evaluating the intrinsic orbital Hall effect (OHE) and spin Hall effect (SHE) in elemental transition metals: fcc Pt, fcc Ti, fcc Cu, fcc Ni, and bcc W, with lattice parameters of 3.92 \AA, 4.10 \AA, 3.61 \AA, 3.52 \AA~and 3.16 \AA, respectively. These calculations are performed using the real-space Kubo-Bastin formalism applied to large supercells, allowing us to access the intrinsic transverse responses without requiring Bloch-state representations. Among these materials, Ni is the only ferromagnet, enabling a comparison between time-reversal symmetric and broken-symmetry systems. 

An important aspect of real-space calculations is the treatment of boundary conditions. Due to the ill-defined nature of the position operator in periodic systems, the computed orbital Hall conductivities can depend on whether periodic or open boundary conditions are used. This effect is particularly relevant for the OHE, where subtle differences arise in the evaluation of the orbital current operator. To ensure consistency with previous results obtained using momentum-space formulations \cite{Go2024}, we adopt open boundary conditions for the final OHE calculations. 

\begin{figure}[h]
    \includegraphics[width=1\columnwidth]{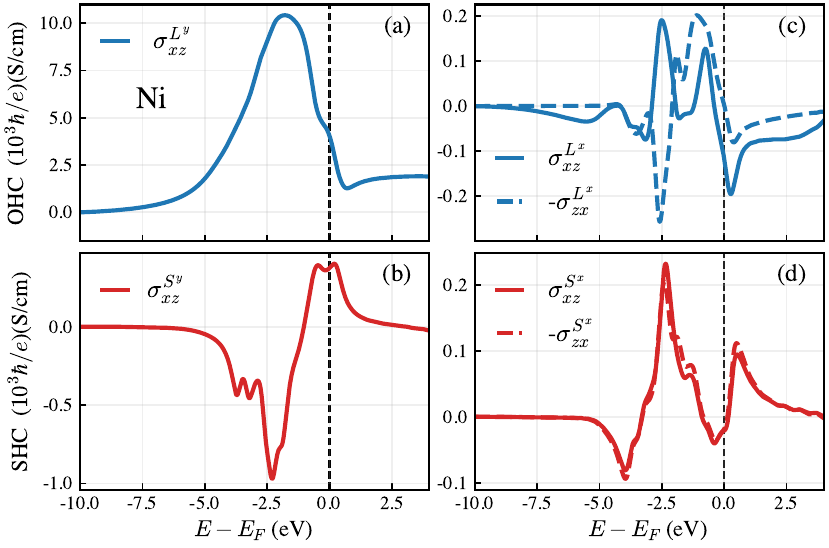}
    
    \caption{Spin and orbital Hall conductivities in ferromagnetic Ni as a function of Fermi energy. 
    Panel (a) shows the orbital Hall conductivity \( \sigma^{L_y}_{xz} \) (blue), and panel (c) the orbital Hall conductivity \( \sigma^{L_x}_{xz} \) (blue). 
    Panels (b) and (d) present the corresponding spin Hall conductivities, \( \sigma^{S_y}_{xz} \) and \( \sigma^{S_x}_{xz} \), respectively (red). 
    The superscripts indicate the angular momentum component (orbital or spin), while the subscripts denote the direction of current flow (\( z \)) and the direction of the applied electric field (\( x \)). 
    Finite values in panels (c) and (d) reflect anomalous contributions enabled by the magnetization of Ni.}
    \label{fig:condNi}
   
\end{figure}
The computed spin and orbital Hall conductivities as functions of the Fermi energy serve as reference values for bulk transport behavior and establish a foundation for understanding current-induced responses in multilayer structures, which will be discussed in the following sections.

In Fig.~\ref{fig:condnm}, we present the calculated orbital Hall conductivity (OHC) and spin Hall conductivity (SHC) for Cu, Ti, Pt, and W as a function of energy. The first two rows show the total OHC and SHC, while the lower two rows provide their orbital-resolved decomposition into $s$-, $p$-, and $d$-orbital contributions. For all materials, we find that the $d$ orbitals dominate the response near the Fermi level, especially in Pt and W, where strong spin-orbit coupling and complex $d$-band structure lead to sizable Hall conductivities. In contrast, Cu displays weaker overall response, with smaller but non-negligible $d$ contributions, while Ti shows a broader distribution of contributions across orbital channels and negligible SHC, as expected, given its weak SOC.

The orbital decomposition provides insight into the microscopic origin of the response, as resolving the signal by orbital character allows us to track the angular momentum content of the contributing states. This is particularly useful in transition metals with mixed $s$-$p$-$d$ hybridization. In W, for instance, the dominant $d$-orbital contribution to both OHC and SHC persists across a wide energy range, while in Ti and Cu, the $p$-orbital contribution is relatively more pronounced.

Our results are consistent with previous studies based on plane-wave DFT methods and Wannier interpolation. In particular, we find good agreement with the spectral features and overall magnitude of the OHC and SHC reported by Salemi \textit{et al.}~\cite{Salemi2022} and of Go \textit{et al.}~\cite{Go2024}.  Furthermore, when comparing with the recent work of Go \textit{et al.}~\cite{Go2024}, we find that our $d$-orbital-resolved OHC curves exhibit a very similar energy dependence to their total OHC, even though they do not perform orbital decomposition. This correspondence reinforces the conclusion that $d$ orbitals dominate the orbital Hall response in heavy transition metals. The quantitative differences between the methods likely stem from the choice of DFT basis and projection schemes, but the key physical trends appear robust across different computational methods.

In Fig.~\ref{fig:condNi}, we show the calculated spin and orbital Hall conductivities for ferromagnetic Ni. The panels are arranged as follows: the first row presents the OHC, while the second row displays the SHC. In each row, we report two different tensor components: the left panels correspond to \( \sigma^{\hat{\mathcal{O}}_y}_{xz} \), and the right panels to \( \sigma^{\hat{\mathcal{O}}_x}_{xz} \). Here, the subscript denotes the spatial geometry — the direction of the response current (\( z \)) and the applied electric field (\( x \)). 

In non-magnetic systems, only time-reversal even terms survive \(( \sigma^{\hat{\mathcal{O}}_y}_{xz} \) in this case), but in ferromagnetic systems like Ni, time-reversal symmetry is broken by the magnetization, enabling the appearance of additional, odd (or anomalous) terms in the conductivity tensor (\( \sigma^{\hat{\mathcal{O}}_x}_{xz} \)). These odd contributions change sign with the magnetization direction and are essential for understanding spintronic and orbitronic responses in magnetic systems.

Our results show finite values for both  \( \sigma^{\hat{\mathcal{O}}_x}_{xz} \)  in the spin and orbital channels, clearly signaling the presence of anomalous spin and orbital Hall effects. These observations are consistent with the symmetry analysis and first-principles calculations presented by Salemi \textit{et al.}~\cite{Salemi2022b}, who decomposed the SHC and OHC into even and odd contributions. In particular, their Figs.~1 and 3 show the spectral behavior of the even and odd conductivities for Ni, which match the trends observed in our calculations, especially the features near the Fermi level.

Additionally, the comparison between  \( \sigma^{\hat{\mathcal{O}}_x}_{xz} \) and  \( -\sigma^{\hat{\mathcal{O}}_x}_{zx} \) in our data provides further insight and it shows the anisotropic behavior of such conductivity component, which is more pronounced on OHC. While \( \sigma^{\hat{\mathcal{O}}_x}_{xz} \) vanishes in the non-magnetic limit, it becomes finite in the ferromagnetic phase and differs in magnitude and sign from \( \sigma^{\hat{\mathcal{O}}_y}_{xz} \). These results confirms that our real-space methodology correctly captures magnetization-induced symmetry breaking and its impact on spin and orbital transport in ferromagnetic materials like Ni.

\section{Spin and Orbital Accumulations in Multilayers}
\begin{figure*}[ht]
    \includegraphics[width=1.5\columnwidth]{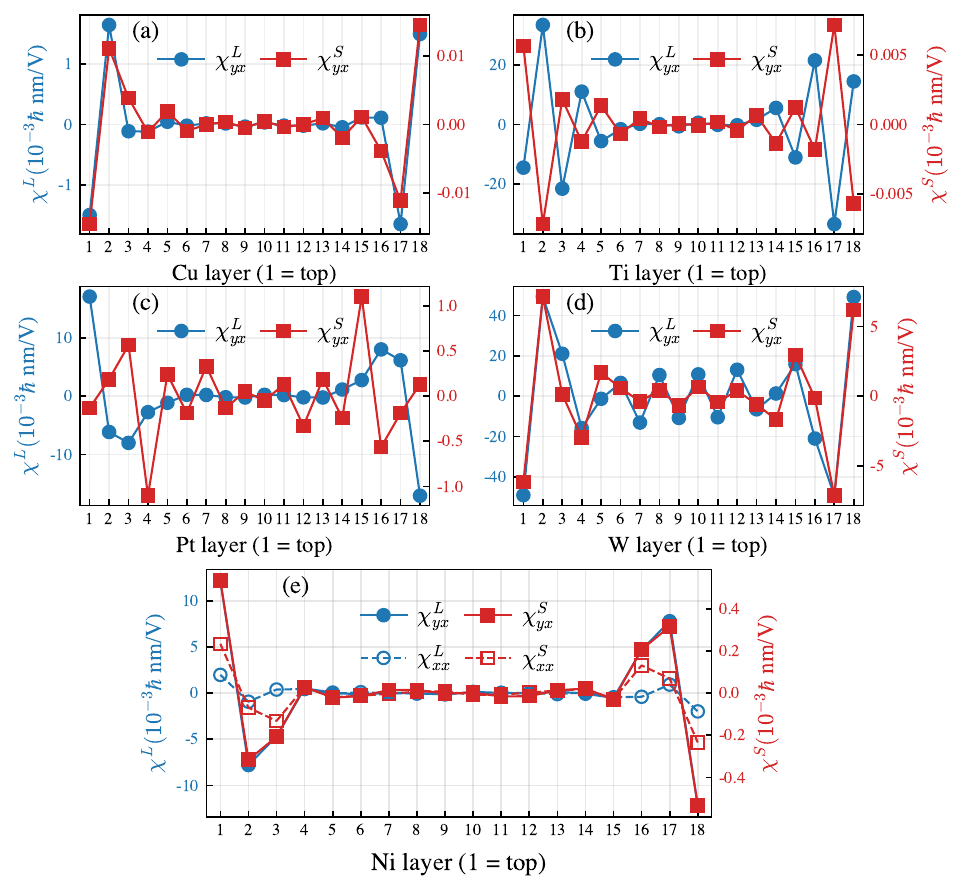}
    \caption{Layer-resolved ME susceptibilities $\chi_{j\beta}^{\hat{\mathbf{O}}}$ related to spin and orbital accumulations in 18-layer slabs of transition metals: (a) Cu, (b) Ti, (c) Pt, (d) W, and (e) Ni. Each panel displays the spatial profiles of spin (red squares) and orbital (blue circles) accumulations along the out-of-plane direction (layer index: top = 1). The left vertical axis (blue) refers to orbital accumulation components $\chi^{L}_{xx}$ and/or $\chi^{L}_{yx}$, and the right vertical axis (red) corresponds to spin accumulation components $\chi^{S}_{xx}$ and/or $\chi^{S}_{yx}$. Accumulations are concentrated near the surfaces and decay inside the slab, with material-dependent amplitude and spatial profile.}
    \label{fig:accum}
\end{figure*}
To investigate the spatial profile of spin and orbital angular momentum accumulation, we consider finite multilayer geometries. The electric field is applied along the $x$-direction, and we evaluate the induced spin and orbital moments at each atomic layer using the real-space Kubo-Bastin formalism that captures both interband and intraband contributions to the spin and orbital angular momentum accumulation. 

We begin with a reference system composed of a single transition-metal slab consisting of 18 atomic layers, and compute the non-equilibrium spin and orbital accumulations across the structure. This geometry allows us to probe the intrinsic layer-resolved response. The surface direction chosen was the (001) in order to illustrate the spin and orbital accumulation, $\delta S_y$ and $\delta L_y$, respectively, across the layers stacked along the z-direction.

Next, using the same geometry as before, we consider heterostructures composed of two atomic layers of a ferromagnetic material (Ni, Co, or Fe) in contact with a nonmagnetic transition metal, forming a multilayer with a total of 11 layers: 2 layers of the ferromagnet and 9 layers of the nonmagnetic TM. In this case, the ferromagnetic layers are assumed to sit on the nonmagnetic stack (fcc for Ti and bcc for W) and inherit its lattice parameter. This geometry enables us to study the influence of magnetic proximity and interfacial scattering on angular momentum accumulation, and allows for a direct comparison between different ferromagnetic elements. The spatial resolution of our method provides a clear layer-by-layer view of the build-up of spin and orbital accumulation near the interface, which is crucial for understanding the generation of current-induced torques in spintronic and orbitronic devices.

Figure~\ref{fig:accum} shows the layer-resolved spin and orbital angular momentum accumulation in slabs of Cu, Ti, Pt, Ni, and W, each consisting of 18 atomic layers and terminated by vacuum on both sides. The electric field is applied in the $x$-direction and the response is computed directly in real space. For each element, the induced spin and orbital ME susceptibility $\chi_{yx}^{\hat{\mathcal{O}}}$ are plotted as a function of the layer index, revealing the symmetry and spatial extent of the angular momentum accumulation.

In nonmagnetic materials such as Cu, Ti, Pt, and W, the spin and orbital accumulations are antisymmetric, as one should expect for the spin and orbital Hall effects, and it is strongest at vicinity of the surfaces, decaying toward the center of the slab. 
The orbital ME susceptibility, $\chi_{yx}^{L}$, is generally much larger in magnitude than the spin ME susceptibility, $\chi_{yx}^{S}$, with both quantities plotted on separate vertical axes for clarity.

However, one can clearly see the differences between the accumulations across the different TMs. For a improved physical insight, from now on we will refer to the spin and orbital accumulation related to $\chi_{i\beta}^{\hat{\mathcal{O}}}$. We now compare the spin and orbital accumulation profiles obtained for 18-layer stacks of different transition metals:   

In Cu (Fig.~\ref{fig:accum}a), the accumulation magnitudes are small and decay quickly toward the interior of the sample. Ti (Fig.~\ref{fig:accum}b) displays stronger orbital accumulation at the surfaces and exhibits more significant layer-to-layer fluctuations, while the spin component remains very small and presents opposite sign. In Pt (Fig.~\ref{fig:accum}c), both spin and orbital accumulation are present, with the spin component showing more pronounced oscillations inside the slab compared to the orbital one. The W sample (Fig.~\ref{fig:accum}d) shows the largest accumulation amplitudes among the non-magnetic materials, especially in the orbital channel, not only with sizable surface accumulation but also non-negligible values extending into the bulk. Its orbital and spin accumulations have the same sign, contrary to what occurs for the SHC and OHC. 
\begin{figure}[h]
    \includegraphics[width=1\columnwidth]{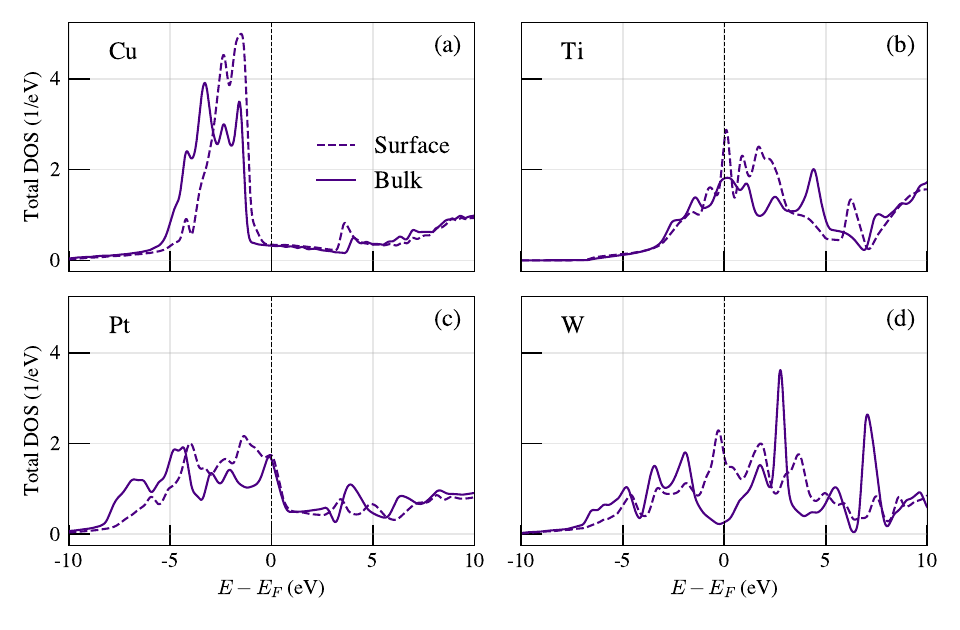}
        \caption{Comparison between the total density of states (DOS) at the surface (dashed lines) and in the bulk (solid lines) for four transition metals: (a) Cu, (b) Ti, (c) Pt, and (d) W. The horizontal axis shows the energy relative to the Fermi level \( E - E_F \), and the vertical axis shows the DOS in arbitrary units. In some cases, a clear deviation is observed between surface and bulk DOS near the Fermi level.}
   
    \label{fig:dosTM}
\end{figure}

While Hall conductivities are intrinsic bulk quantities, the accumulations occur at the boundaries and depend not only on the magnitude of the SHC and OHC but also on the surface electronic structure. For instance, in W, we find that the relative signs of the spin and orbital accumulations differ from those of the corresponding Hall conductivities, indicating that interfacial or surface effects can play a decisive role. A possible explanation lies in the difference between the density of states (DOS) at the surface and in the interior of the material. Since the spin and orbital accumulations arise from a non-equilibrium population of carriers, it is sensitive to the available states near the Fermi level.

To better understand the differences in sign and magnitude between the bulk spin and orbital Hall conductivities and the corresponding surface accumulations, we examine the total density of states (DOS) at the surface and in the bulk of the transition metals, shown in Fig.~\ref{fig:dosTM}. This figure compares the total DOS projected onto the top layer (dashed lines) and onto a central layer (solid lines) for each material.
The data reveal that surface states can strongly influence the local electronic structure. For W, in particular, there is a considerable difference between the bulk and surface DOS near the Fermi level, with the surface displaying a pronounced peak. This enhancement is consistent with the large orbital and spin accumulations observed in Fig.~\ref{fig:accum}d and may also explain the relative sign difference with respect to the bulk OHC and SHC. These observations suggest that the surface electronic structure plays a crucial role in determining the sign and magnitude of the induced accumulations. The interplay between surface DOS and the layer-resolved responses highlights the importance of going beyond bulk properties when interpreting spin and orbital accumulation in confined geometries.

 In the ferromagnetic Ni slab, we observe qualitatively different behavior compared to nonmagnetic materials. We detect non-negligible components of both $\chi_{xx}^{S}$ and $\chi_{xx}^{L}$ in addition to the $\chi_{yx}^{\hat{\mathcal{O}}}$, reflecting the broken time-reversal symmetry inherent to ferromagnets. These observations are consistent with the presence of magnetic counterparts to the conventional Hall effects: the so-called magnetic spin Hall effect (MSHE) and magnetic orbital Hall effect (MOHE). First analyzed in bulk Fe, Co, and Ni by Salemi and Oppeneer \cite{Salemi2022}, the MSHE and MOHE represent time-reversal odd contributions that coexist with the conventional, time-reversal even SHE and OHE. In particular, MSHE in Ni has been shown to be significant and may even reverse sign relative to the SHE, while MOHE, although generally smaller than OHE, still contributes appreciably to angular momentum accumulation \cite{Salemi2022}. Our layer-resolved results align with these theoretical predictions by showing that ferromagnetism induces anomalous accumulation and activates in-plane components of angular momentum that are absent in nonmagnetic cases.
\begin{figure*}
    \centering
    \includegraphics[width=2.0\columnwidth]{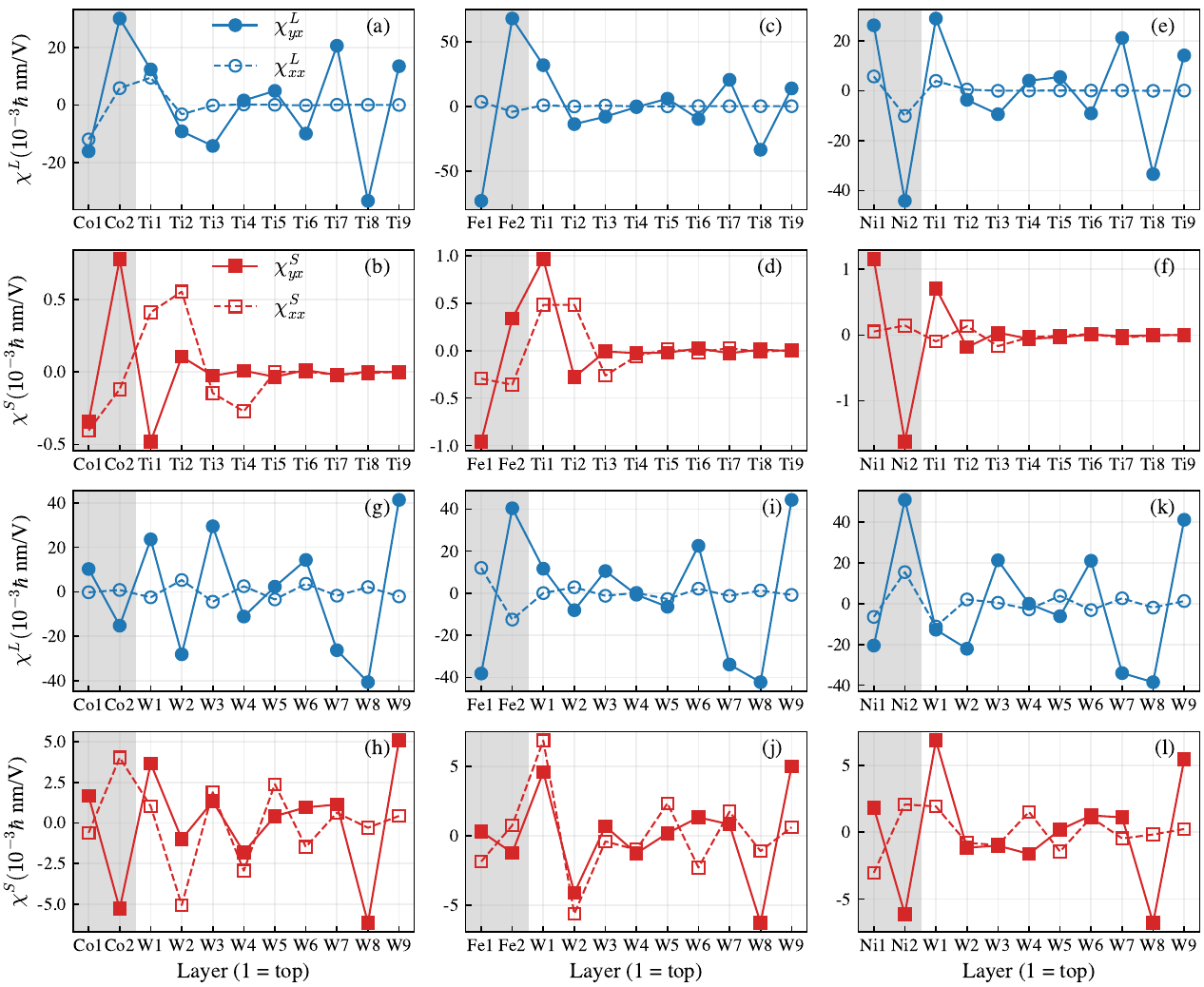}
    
    \caption{%
        Layer-resolved orbital (top panel, circles) and spin (bottom panel, squares) accumulations for heterostructures composed of two layers of a ferromagnet followed by nine layers of a transition metal: 
        (a)-(b) Co/Ti, 
        (c)-(d) Fe/Ti, 
        (e)-(f) Ni/Ti, 
        (g)-(h) Co/W, 
        (i)-(j) Fe/W, and 
        (k)-(l) Ni/W.
        Each panel shows the in-plane components \(L_x\), \(L_y\) (orbital, in blue) or \(S_x\), \(S_y\) (spin, in red) as a function of the layer index.
    }
    \label{fig:accml}
\end{figure*}
 
We now analyze the spin and orbital accumulation profiles in heterostructures composed of two layers of a ferromagnet (Fe, Co, or Ni) followed by nine layers of a transition metal (Ti or W). Figure~\ref{fig:accml} presents the results for all six combinations, with separate panels showing the in-plane components of orbital ($\chi_{xx}^{L}$, $\chi_{yx}^{L}$) and spin ($\chi_{xx}^{S}$, $\chi_{yx}^{S}$) ME susceptibilities.

In all systems, the largest accumulations related to $\chi_{j\beta}^{\hat{\mathcal{O}}}$ appear near the interface, highlighting the role of symmetry breaking and interfacial hybridization in generating angular momentum responses. We observe that Fe-based heterostructures exhibit the strongest orbital accumulation, particularly in the Fe/Ti system. Co also displays large orbital accumulations, comparable to or even larger than those of Ni. In contrast, the orbital accumulation in Ni-based systems, although sizable, is systematically smaller than in Fe and Co, especially in the Ni/Ti case.

The spin accumulation, on the other hand, is generally more localized and of smaller amplitude than the orbital counterpart, confirming the more delocalized nature of orbital angular momentum. This trend is particularly evident in the W-based heterostructures, where the orbital accumulation extends several layers into the transition metal and displays smoother spatial variation. Overall, these results show that both the amplitude and spatial extent of angular momentum accumulation depend sensitively on the choice of ferromagnet and adjacent transition metal and have direct implications for the resulting spin-orbit torques in each system.

While the accumulation profiles presented above offer valuable insight into the behavior of angular momentum near surfaces and interfaces, they do not directly correspond to the spin or orbital torques acting on the ferromagnet. In particular, the torque is not proportional to the accumulated angular momentum but is instead governed by the divergence of the angular momentum flux \cite{Go2020a,Go2020b}. This distinction is important in the context of spin-orbit torques at TM/FM interfaces. While a sizable spin or orbital accumulation at the interface may suggest the presence of a torque, the actual transfer of angular momentum depends on the flux of these quantities from the adjacent non-magnetic metal into the ferromagnet. Since our calculations do not yet include a reliable estimate of the flux contributions, the accumulation results should be interpreted as indirect indicators of the interfacial torque, not as a direct measure. Future extensions of this work including layer-resolved flux calculations will enable a more quantitative connection between bulk responses, interfacial accumulations, and the resulting torques.

\section{Conclusion}

We have developed and applied a real-space, linear-scaling methodology to compute orbital and spin Hall conductivities as well as their corresponding layer-resolved accumulations in metallic multilayers and heterostructures. The method, based on the RS-LMTO-ASA approach combined with a Chebyshev expansion of the Green’s functions, allows for the calculation of linear response transport properties directly in real space. This avoids the need for projections onto Wannier functions or the use of periodic boundary conditions, making it ideally suited for systems with broken translational symmetry, such as surfaces, interfaces, and disordered structures.

Our benchmark calculations for bulk spin and orbital Hall conductivities in Cu, Ti, Pt, W, and Ni reproduce well-established trends and show excellent agreement with previous studies based on Bloch-based first-principles methods, such as those by Go \textit{et al.}\cite{Go2024} and Salemi \textit{et al.}\cite{Salemi2022}. However, the power of the real-space approach is best illustrated in its ability to capture spatially resolved accumulations across finite slabs, where surface and interfacial effects become crucial. We have shown that the relation between bulk conductivities and surface accumulations is not straightforward: in some cases, notably W, the surface accumulations exhibit opposite signs to the bulk Hall conductivities. 

To shed light on this discrepancy, we computed the density of states projected on the surface and the bulk layers of the transition-metal slabs. The data reveal that modifications of the spectral weight near the Fermi level at the surface can lead to changes in the direction and magnitude of the induced accumulations. This highlights the importance of surface electronic structure and its deviation from bulk behavior in determining the actual response of confined systems.

In ferromagnetic Ni, we observed finite values for both \( \sigma_{xz}^{S_x} \), \( \sigma_{xz}^{S_y} \) and their orbital counterparts, in line with recent decompositions of the spin and orbital Hall conductivities into time-reversal even and odd components \cite{Salemi2022b}. These findings confirm that the real-space formalism captures symmetry-breaking effects due to magnetization and is well suited for studying transport phenomena in magnetic materials.

Altogether, this work demonstrates that the real-space methodology developed here is a powerful and scalable tool for investigating orbital and spin transport beyond the bulk limit. It enables direct access to local responses and provides a pathway to study systems with reduced symmetry, disorder, and complex interfaces—scenarios that are increasingly relevant in the design of next-generation spintronic and orbitronic devices.

\section{Acknowledgments}
R.C.\ acknowledges insightful discussions with Danny Thonig and Roberto Bechara Muniz. T.G.R.\ acknowledges support from the EIC Pathfinder OPEN grant 101129641 “OBELIX” and the Brazilian agency CNPq and insightful discussions with Dongwook Go. 
H.U.R.S.\ acknowledges financial support from the Swedish Research Council (Vetenskapsrådet, VR) grant number 2024-04652 and funding
from the European Research Council (ERC) under the
European Union’s Horizon 2020 research and innovation
programme (grant agreement No. 854843-FASTCORR).
A.B. acknowledges support from eSSENCE, the Carl Trygger foundation and the AI4Research center at Uppsala University. A.B.K. acknowledges support from CAPES and CNPq, Brazil; the INCT of Materials Informatics and the INCT of Spintronics and Advanced Magnetic Nanostructures, CNPq, Brazil.
\newpage
\bibliography{references_tgr}
\appendix

\end{document}